\documentclass[pre,aps,twocolumn,floatfix,showpacs]{revtex4-1}

\usepackage{graphicx}
\usepackage{amsmath}
\usepackage{amssymb}
\usepackage{wasysym}

\newcommand{\bra}[1]{\langle #1|}
\newcommand{\ket}[1]{|#1\rangle}
\newcommand{\expectation}[1]{\left\langle #1\right\rangle}

\newcommand{\Tr}{\text{Tr}}

\newcommand{\gammaext}{\gamma_\text{ext}}
\newcommand{\gammaSSET}{\gamma_\text{SSET}}
\newcommand{\xfp}{x_\text{fp}}
\newcommand{\Vds}{V_\text{ds}}

\begin{document}

\title{Phase Transitions in Trajectories of a Superconducting Single-Electron Transistor Coupled to a Resonator}

\author{Sam Genway}
\author{Juan P. Garrahan}
\author{Igor Lesanovsky}
\author{Andrew D. Armour}
\affiliation{School
of Physics and Astronomy, The University of Nottingham, Nottingham,
NG7 2RD, United Kingdom}


\date{\today}

\begin{abstract} 
Recent progress in the study of dynamical phase transitions has been made with a large-deviation approach to study trajectories of stochastic jumps using a thermodynamic formalism.  We study this method applied to an open quantum system consisting of a superconducting single-electron transistor, near the Josephson quasiparticle resonance, coupled to a resonator.  We find that the dynamical behavior shown in rare trajectories can be rich even when the mean dynamical activity is small and thus the formalism gives insights into the form of fluctuations.
 The structure of the dynamical phase diagram found from the quantum-jump trajectories of the resonator is studied and we see that sharp transitions in the dynamical activity may be related to the appearance and disappearance of bistabilities in the state of the resonator as system parameters are changed.  We also demonstrate that for a fast resonator, the trajectories of quasiparticles are similar to the resonator trajectories.
\end{abstract}

\maketitle

\section{Introduction}

Transitions in the dynamical behavior of open quantum systems may be seen for a variety of driven systems.  Notable examples include the laser~\cite{Scully1997}, atoms coupled to optical cavity modes such as in the micromaser~\cite{Walther2006,Haroche2006} and pumped Bose-Einstein condensates~\cite{Baumann2010}, mechanical resonators coupled to optical cavities~\cite{Marquardt2006}, and nanoelectromechanical systems~\cite{Rodrigues2007}.  Recently, an alternative perspective on the dynamics of open quantum systems has been developed~\cite{Garrahan2010,Garrahan2011} by studying ensembles of quantum-jump trajectories.  Applying the so-called \emph{s-ensemble} \cite{Lecomte:2007fk,Garrahan:2007la,Hedges:2009fk}
has given insights into dynamical crossovers by extending the space of parameters~\cite{Garrahan2010,Garrahan2011,Li2011,Ates2011}.  The new parameter `$s$' is a counting field which may be adjusted to give ensembles of trajectories biased towards fewer or more quantum jumps.  Varying $s$ allows the phase structure of quantum-jump trajectories to be explored in a way analogous to tuning parameters across an equilibrium phase transition, such as the temperature in a ferromagnetic transition.

The $s$-ensemble is developed in a formalism using a large-deviation method~\cite{Touchette2009}, which treats the statistics of dynamical trajectories in the same way equilibrium statistical mechanics treats statistical ensembles of configurations \cite{Lecomte:2007fk,Garrahan:2007la,Hedges:2009fk}.  Just as order parameters, such as the magnetization for a ferromagnetic transition, are used to characterize configurations, we can use a dynamical order parameter to characterize quantum-jump trajectories.  The number of quantum jumps $K$ in a given time $t$ is the extensive conjugate variable to the counting field $s$.  We therefore consider the \emph{activity} $k$, defined as the number of quantum jumps in unit time $K/t$, as a dynamical order parameter.  Characterizing dynamical phases in this way has been shown to give rich phase diagrams with phase boundaries marked by discontinuities in $k$, or its derivative with respect to system parameters, occurring at first and second-order transitions~\cite{Garrahan2011}.  For the unbiased physical dynamics when $s=0$, the transitions typically appear rounded and we understand these to be crossovers in dynamical behavior, which become transitions in an appropriate thermodynamic limit~\footnote{
Since we are dealing with trajectories, the thermodynamic limit corresponds to taking both the observation time and the system size (i.e. the dimension of Hilbert space) to infinity.  For example, for the latter in the case of quantum harmonic oscillators studied here, this limit is found for large $n$.}.

Physical dynamics correspond to the $s=0$ ensemble of trajectories.  The method gives a new perspective on these unbiased $s=0$ dynamics through the structure of phase diagrams parametrized by $s$.  Furthermore, it should be noted that the statistics of trajectories when $s \ne 0$ can be inferred from accurate measurements of the physical trajectories.  The $s$-ensemble has already shown interesting results for driven 2- and 3-level dissipative systems.  For example, it has been shown that for a particular ratio of driving and dissipation~\cite{Garrahan2010}, the two-level system exhibits a surprising self-similarity in the fluctuation properties for all $s$, so that all trajectories look the same if rescaled by the number of counts, $k$.  For the three-level system studied in~\cite{Garrahan2010} it was shown that the intermittent bursts of quantum jumps~\cite{Plenio1998} in the dynamics arise from a sharp crossover in activity occurring at $s=0$.   In Ref.~\cite{Garrahan2011}, dynamical phases of the micromaser were studied.  The micromaser consists of a series of two-level atoms passing through a microwave cavity in their excited state.  As the flux of atoms is increased, the atoms drive the cavity through different dynamical states.  It was shown that, as a function of the counting field $s$ and the atom flux, distinct dynamical phases are seen, characterized by different values of the activity.

In this paper, we examine a more complex system consisting of two coupled components with non-trivial internal dynamics.  We study a system constructed by coupling a resonator to a superconducting single-electron transistor (SSET) close to a Josephson quasiparticle resonance (JQP)~\cite{Blencowe2004, Blencowe2005, Clerk2002, Clerk2005, LaHaye2004, Choi2001, Naik2006, Astafiev2007}.  
Both the resonator and SSET are themselves open quantum systems because the resonator loses energy to its surroundings and Cooper pairs on the SSET break into quasiparticles which, one-by-one, decay incoherently off the SSET into a second lead.  This allows us alternative ways of generating trajectories;  we may either keep a time record of quanta emitted by the resonator or measure charges leaving the SSET island.  While SSET-resonator systems were first demonstrated with nanomechanical beam resonators~\cite{Naik2006}, superconducting stripline resonators have also been used in more recent experiments~\cite{Astafiev2007, Frey2012}.  It is the latter set up which, along with very recent progress towards detecting single microwave photons from stripline resonators~\cite{Bozyigit2011,Eichler2011,Zakka-Bajjani2010,Chen2011}, motivates our study of quantum jump trajectories presented here.

This paper is organized as follows.  In section~\ref{Model}, we introduce the open-system model for the SSET and resonator before describing how the trajectories of the system may be studied using the $s$-ensemble.  In section~\ref{Exact}, we present numerical results for the trajectories of quantum jumps in the resonator.  Then, in section~\ref{Mean}, we discuss a mean-field theory which describes the resonator limit cycles at $s=0$.  From this we develop an effective stochastic master equation for the resonator, from which we derive the activity of the resonator for all values of $s$.   Finally, in section~\ref{Quasi} we demonstrate briefly that studying the trajectories of quasiparticles allows the resonator trajectories to be inferred in certain regimes.  In section~\ref{Conclusions} we state our conclusions.

\section{Model and Formalism}
\label{Model}

\begin{figure}
\includegraphics[scale=1.0]{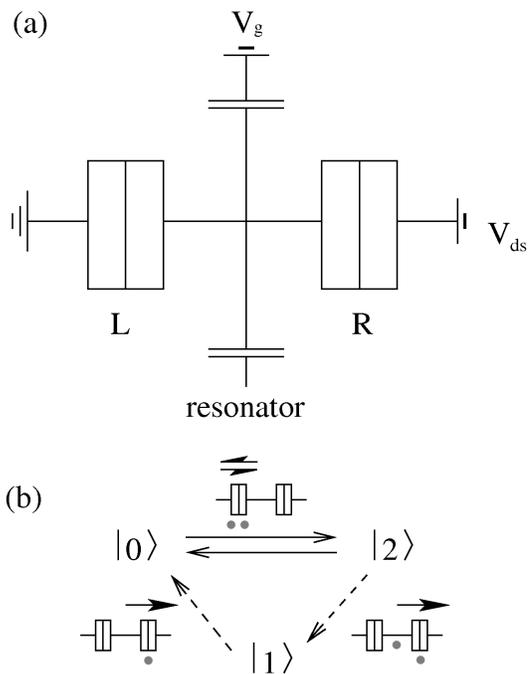}
\caption{(a) Schematic diagram of the SSET-resonator system showing left and right Josephson junctions labeled `L' and `R', with gate voltage $V_\text{g}$ applied across a gate capacitor and drain-source voltage $\Vds$ across both Josephson junctions.  The resonator is also capacitively coupled to the SSET island.  (b) Illustration of the JQP cycle.  Cooper-pair tunneling occurs at the left-hand junction between states $\ket{0}$ and $\ket{2}$.  This is interrupted by the quasiparticle decay to state $\ket{1}$ which in turn decays back to state $\ket{0}$.}
\label{schematic}
\end{figure}

The tandem of a resonator weakly coupled to an SSET near the JQP resonance may be explored with an open quantum system description~\cite{Rodrigues2007}.  The SSET island consists of a left- and a right-hand Josephson junction, a gate capacitor across which a gate voltage is applied and a capacitively-coupled resonator, shown in Fig.~\ref{schematic}(a).  The detuning from the JQP resonance in the left-hand Josephson junction can be controlled with drain-source and gate voltages, $\Vds$ and $V_g$, while a resonator gate voltage also allows the strength of the coupling to the resonator to be adjusted.  Near the JQP resonance, Cooper pairs tunnel coherently across a Josephson junction, causing oscillations between SSET island states $\ket{0}$ and $\ket{2}$, with zero and two extra charges on the SSET respectively.  However, the incoherent tunneling of quasiparticles through the right-hand lead takes the SSET island to the state $\ket{1}$, with a single extra charge, and then back to the state $\ket{0}$, as sketched in Fig.~\ref{schematic}(b).  The coupling of a resonator to the SSET island allows for energy exchange between resonator and SSET.  Depending on the sign of the detuning from the JQP resonance, the resonator is either driven into states of self-sustaining oscillation~\cite{Astafiev2007}, or experiences a cooling effect~\cite{Naik2006} because of its coupling to the SSET.

  In the open-system model, the coherent oscillations of Cooper pairs across the left-hand Josephson junction, the motion of the resonator and the coupling of the resonator position to the charge on the SSET island are all included in an effective Hamiltonian.  While the resonator may be a stripline resonator~\cite{Astafiev2007} or a nanomechanical beam~\cite{Naik2006}, for clarity we will stick to language more appropriate for a mechanical oscillator.  Both the dissipative effects of the environment on the resonator and the decay of quasiparticles from the SSET island and the right-hand lead can be described by Lindblad operators such that the dynamics of the SSET-resonator system may be found from a master equation~\cite{Rodrigues2007,Rodrigues2007a}.

\subsection{Master Equation}

We will focus on the regime where the SSET island exhibits coherent oscillations of Cooper pairs across the left-hand junction between charge states of the SSET island $\ket{0}$ and $\ket{2}$.  Including the quasiparticle decay to the right-hand lead requires the inclusion of a third island state $\ket{1}$ with just a single quasiparticle.  Treating the resonator as a single-mode harmonic oscillator with frequency $\omega$ and considering the fluctuations in the resonator displacement to be small compared with the SSET-resonator distance, allows the coupling to be well approximated by expansion to linear order in the oscillator displacement.  The full Hamiltonian for the coherent dynamics is then~\cite{Rodrigues2007,Rodrigues2007a}.
\begin{eqnarray}
H = &\Delta E \ket{2}\bra{2} - \frac{E_J}{2} \left(\ket{0}\bra{2} + \ket{2}\bra{0}\right) + \hbar \omega a^{\dagger} a \nonumber \\
& + C x_s (a^{\dagger} + a)(\ket{1}\bra{1} + 2\ket{2}\bra{2})\,.
\label{eq:ham}
\end{eqnarray}
The difference in electrostatic energy between states $\ket{0}$ and $\ket{2}$ is given by $\Delta E$ and $E_J$ is the Josephson energy of the junctions.  The displacement $x_s$ is the change in the equilibrium position of the resonator when one charge is added to the SSET island, with $C =({\hbar \omega^3 m}/2)^{1/2}$.  In addition to the coherent dynamics governed by the Hamiltonian, dissipation occurs due to both quasiparticle tunneling off the SSET island and the coupling of the resonator to its environment.  Including these effects, the full master equation reads
\begin{equation}
\dot{\rho} = \mathcal{W}(\rho) = -\frac{i}{\hbar} [H,\rho] + \mathcal{L}_{\text{q.p.}}(\rho) + \mathcal{L}_{\text{osc.}}(\rho)
\label{eq:W}
\end{equation}
where $\rho$ is the density matrix for the combined SSET and resonator system.  The quasiparticle decay from states $\ket{2}$ to $\ket{1}$ and from $\ket{1}$ to $\ket{0}$ is described by
\begin{eqnarray}
\mathcal{L}_{\text{q.p.}}(\rho) = &\Gamma(\ket{1}\bra{2}\rho\ket{2}\bra{1} + \ket{0}\bra{1}\rho\ket{1}\bra{0}) \nonumber \\
&- \frac{\Gamma}{2}\left(\{\ket{2}\bra{2},\rho \} + \{\ket{1}\bra{1},\rho \}\right)
\label{eq:Lqp}
\end{eqnarray}
where $\{\cdot,\cdot\}$ denotes the anticommutator and $\Gamma$ is the quasiparticle decay rate which is, for simplicity, taken to be the same for the two processes $\ket{2} \rightarrow \ket{1}$ and $\ket{1} \rightarrow \ket{0}$.  The environment of the resonator will also be included via
\begin{equation}
\mathcal{L}_\text{osc.}(\rho) = \gammaext a \rho a^{\dagger} - \frac{\gammaext}{2}\, \{a^{\dagger}a,\rho\}
\label{eq:Losc}
\end{equation}
where $\gammaext$ is the rate, per energy quantum $\hbar \omega$, at which the resonator loses energy to its environment.  The Lindblad form~\eqref{eq:Losc} describes a thermal bath at zero temperature which, although hard to achieve with a mechanical resonator, is valid for the higher-frequency stripline resonators where $k_B T \ll \hbar \omega$.  Indeed, Eq.~\eqref{eq:W} is valid for a broad range of oscillator frequencies, provided the coupling between the SSET and the resonator is weak.  (A summary of the practical requirements for a system to be well-described by this model is given in the Appendix.)

We assume throughout the model parameters~\cite{Blencowe2005} $\Gamma = V_\text{ds}/eR_J$, with $R_J = h/e^2$ the SSET junction resistance.  We can define the coupling strength in terms of the dimensionless parameter $\lambda$, where $\lambda^2 = m\omega^2 x_s^2/e V_{\text{ds}}$ will be much smaller than unity for the weakly-coupled system studied in this paper \footnote{It may be noted that in several works in the literature, the parameter denoted $\kappa$ is identical to $\lambda^2$ in this paper.}.   

\begin{figure}[htb]
 \begin{centering}
   \includegraphics[scale=1.4]{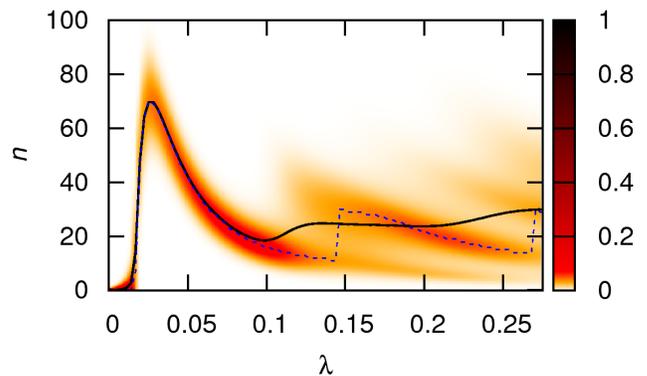}
   \caption{(color online) A density plot showing the state of the resonator in the SSET-resonator system as a function of the coupling strength, $\lambda$, between the resonator and the SSET island.  Shown is the occupation probability of resonator number states $P(n)$ with additional lines showing the expectation $\expectation{n}$ (solid) and the most probable number state $n_\text{mp}$ (dashed).  The system parameters are $\Gamma = V_\text{ds}/eR_J$, with $R_J = h/e^2$, $E_J=e\Vds/16$ and $\gammaext = 0.0005 \Gamma$.}
   \label{EJ1_s0}
 \end{centering}
\end{figure}

A characteristic of this system is the ability of the SSET to drive~\cite{Rodrigues2007,Astafiev2007} or cool~\cite{Clerk2005,Blencowe2005,Naik2006} the resonator depending on the sign of $\Delta E$, which is set by the detuning of the SSET bias point from the JQP resonance.   When $\Delta E<0$, the SSET can drive the resonator into different steady states as the SSET-resonator coupling strength is increased.   When decoupled from the SSET, the resonator remains in its ground state.  Upon increasing the coupling, a continuous crossover is reached where the oscillator enters a state of self-sustaining oscillation.  This is illustrated by the distribution over the number states $P(n)$ for the oscillator in its steady state, shown in Fig.~\ref{EJ1_s0} for the model described by Eq.~\eqref{eq:W}.  It was shown in Ref.~\cite{Rodrigues2007} that the resonator Wigner functions indicate that the crossover is from a fixed-point state in phase space to a limit-cycle state.  These steady states correspond classically to a static state, where the amplitude is zero, and a state with well-defined amplitude undergoing harmonic oscillations, respectively.   Upon increasing the coupling strength further, it is possible to drive the resonator through a series of first-order crossovers.  These dynamical crossovers are associated with changes between limit cycles of different amplitudes.  The first order crossovers are also illustrated in Fig.~\ref{EJ1_s0}, where the most-probable resonator state $n_\text{mp}$ is plotted.  This allows identification of the mid-points of the crossovers, where $n_\text{mp}$ shows sharp jumps.

\subsection{Quantum Trajectories and the $s$-Ensemble}

We now turn to the study of the statistics of quantum-jump trajectories in the system.  A trajectory is a time record of quantum jumps associated with particular stochastic operators in the master equation~\cite{Garrahan2010,Garrahan2011}.  In the SSET-resonator system we have the choice of counting charges leaving the SSET island through the right-hand lead or counting quanta entering the environment of the resonator.   The main focus of this paper is on trajectories of the resonator, but we study briefly charge trajectories in Section~\ref{Quasi}.  By considering the projection of the density matrix $\rho(t)$ on to the subspace where $K$ events have occurred within the time $t$, we can define the reduced density matrix $\rho^{(K)}(t)$ such that the probability to observe $K$ events in time $t$ is given by $P_t(K) = \Tr[\rho^{(K)}(t)]$.  After long times, $P_t(K)$ takes a large-deviation form~\cite{Touchette2009}:
\begin{equation}
P_t(K) \simeq e^{-t\varphi(K/t)}
\label{eq:trajdist}
\end{equation}
where $\varphi(K/t)$ is the large-deviation function which allows a complete description of the statistics of $K$ at long times.  The $s$-ensemble~\cite{Garrahan2010,Garrahan2011} enters by considering a moment-generating function associated with counting probabilities
\begin{equation}
Z_t(s) = \sum_K P_t(K) e^{-sK} \simeq e^{t\theta(s)}
\label{eq:Z}
\end{equation}
where the second equality is valid for long times.  The large deviation functions $\theta(s)$ and $\varphi(k)$, with $k=K/t$, are related by the Legendre transform $\theta(s) = -\min_k[\varphi(k) + ks]$~\cite{Budini2011,Touchette2009}.  The $s$-field is the intensive conjugate field to the time-extensive number of events $K$, and the scaled activity $k$ may be used as an order parameter to distinguish different dynamical phases.  The large deviation functions $\varphi(k)$ and $\theta(s)$ \cite{Lecomte:2007fk,Garrahan:2007la,Hedges:2009fk} themselves take on roles analogous to those of entropy density and free energy density density in equilibrium statistical mechanics.

The large-deviation function $\theta(s)$ will be studied throughout the rest of this paper as it encodes the full distribution of trajectories~\cite{Garrahan2010,Budini2011}.  Furthermore, analogous to minus a free energy in equilibrium statistical mechanics, discontinuities in the derivatives of $\theta(s)$ have been found to correspond to phase transitions in ensembles of trajectories of the dynamics~\cite{Garrahan2010}.  The large-deviation function $\theta(s)$ may be found from a generalized master equation~\cite{Esposito2009}, which takes the form of an $s$-dependent modification~\cite{Garrahan2010} to Eq.~\eqref{eq:W}.  First, we introduce an $s$-biased density matrix, $\rho_s$, defined by
\begin{equation}
\rho_s(t) = \sum_K \rho^{(K)}(t) \, e^{-sK}\,.
\end{equation}
Then, when studying the trajectories associated with photons leaving the resonator, $\theta(s)$  is the largest eigenvalue of the generalized master equation
\begin{equation}
\dot{\rho_s} = \mathcal{W}_s(\rho_s) = -\frac{i}{\hbar} [H,\rho_s] + \mathcal{L}^\text{q.p.}(\rho_s) + \mathcal{L}_s^\text{osc.}(\rho_s)
\label{eq:Ws_osc}
\end{equation}
where
\begin{equation}
\mathcal{L}_s^\text{osc.}(\rho_s) = \gammaext \left(e^{-s} a \rho_s a^{\dagger} - \frac{1}{2} \{a^{\dagger} a,\rho_s\}\right)\,.
\end{equation}
For $s \longrightarrow 0$, the superoperator $\mathcal{W}_s$ collapses to $\mathcal{W}$ and, necessarily, $\theta(s) \longrightarrow 0$ as this corresponds to (the usual trace-preserving) physical dynamics.  Away from $s=0$, the dynamics are biased by the $s$-field towards rare trajectories with, for $s>0$, fewer events $K$ within a time $t$ or, for $s<0$, more events.  We will refer to these as, respectively, \emph{less active} and \emph{more active} rare trajectories.  While the superoperator $\mathcal{W}_s$ is no longer trace-preserving when $s\ne 0$, the mapping is still completely positive~\cite{Catana2011} as it can straightforwardly be shown to be of Kraus form~\cite{Bengtsson2006}.  For all $s$, effective steady-state properties can be determined from the right eigenmatrix of $\mathcal{W}_s$ associated with the largest eigenvalue $\theta(s)$.  Some pedagogical examples of the $s$-ensemble applied to quantum master equations are discussed in Ref.~\cite{Garrahan2010}.

\section{Numerical Results}
\label{Exact}
The SSET-resonator generalized master equation~\eqref{eq:Ws_osc} may be expressed in matrix form and diagonalized numerically, provided the basis for the oscillator is truncated.  For the oscillator damping $\gammaext/\Gamma = 0.0005$ studied in this paper, the Josephson energies will be chosen so that the maximum energy of the resonator $n \hbar \omega$ has $n \apprle 200$~\cite{Harvey2008,Koerting2009}.  However, since the charge on the SSET couples to coherences between the oscillator eigenstates, it is necessary to keep a basis with off-diagonal density matrix elements for the oscillator.  In practice, for the coupling linear in the oscillator position, it is possible to truncate the basis such that only eigenstates which differ in energy by $m\hbar\omega$ (with $m<n$) have coherences preserved.  The choice of $m$ may be tested numerically to ensure that the results are not sensitive to this truncation.  For the SSET island, coherences are only generated between the states $\ket{0}$ and $\ket{2}$, such that propagation of the SSET density matrix in time requires inclusion of just five of the nine matrix elements.  Therefore, the exact numerical study of $\theta(s)$ requires extraction of the largest real eigenvalue of an approximately $5nm\times 5nm$ matrix.  In the results which follow, we implemented an Arnoldi iteration scheme~\cite{Arpack} to find $\theta(s)$ from the generalized master equation.

\subsection{Resonator Dynamics}

When the SSET bias point is chosen such that $\Delta E <0$, energy is transferred from the SSET island to the resonator on average.  When increasing the strength of the coupling between the SSET and the resonator, $\lambda$, the resonator is driven away from its fixed-point state through a continuous crossover into a limit-cycle.  Further increases in $\lambda$ drive the oscillator through a series of apparently first-order dynamical crossovers as the resonator demonstrates a series of bistabilities between different limit cycles, as illustrated in Fig.~\ref{EJ1_s0}.  


By measuring the emission of photons from a stripline resonator~\cite{Bozyigit2011,Eichler2011,Zakka-Bajjani2010,Chen2011}, it would be possible to infer the state of the resonator. Due to the linear coupling to the bath, the activity is proportional to both the energy of the oscillator $\hbar\omega n$ and the decay rate $\gammaext$.    The activity may be extracted in the $s$-ensemble from the large deviation function $\theta(s)$.  From Eq.~\eqref{eq:Z}, it is clear in the steady state that the activity $k = \expectation{K}/t$ may be found as:
\begin{equation}
k= -\frac{1}{t}\frac{\partial}{\partial s}\sum_K P_t(K) e^{-sK} \Big|_{s=0} = -\theta'(0)
\label{eq:act}
\end{equation}
where the prime in the last equality denotes differentiation with respect to $s$.  Higher derivatives of $\theta(s)$ correspond to higher moments of the distribution of photon emissions from the resonator.  

In contrast to studies of full counting statistics~\cite{Flindt2005,Flindt2009} which focus on the $s=0$ physical dynamics (see also Ref.~\cite{Ivanov2010}), we will study the dynamical phases as a function of the counting field $s$.  In this way, the dynamical behavior at $s=0$ as, for example, a function of $\lambda$ may be understood through its proximity to the phase boundaries in the $\lambda-s$ plane.  We will first explore the full dynamical phase diagram for the case of a resonator coupled to an SSET where the quasiparticle decay rate is matched to the oscillator frequency, $\omega = \Gamma$, and the SSET level separation $\Delta E$ is negative, such that there is, on average, transfer of energy from the SSET island to the resonator.  

\begin{figure}[!htb]
 \begin{centering}
   \includegraphics[scale=1.2]{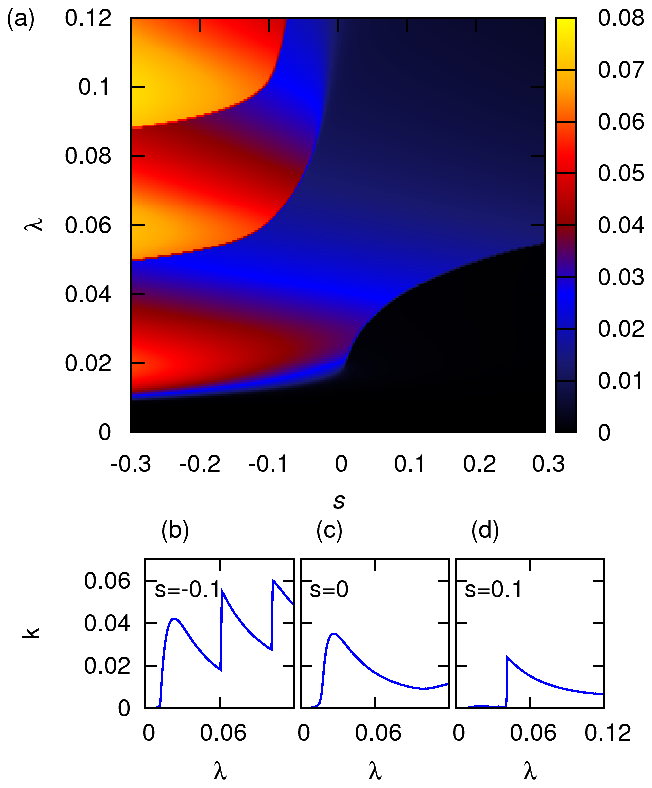}
      \includegraphics[scale=1.2]{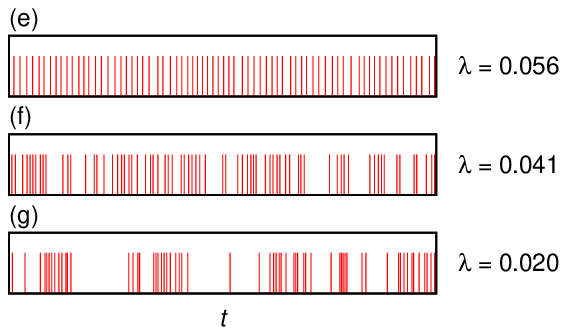}
   \caption{(color online) (a) A plot of the activity $k$ as a function of coupling strength $\lambda$ and counting field $s$ for system parameters as in Fig.~\ref{EJ1_s0}, where $E_J/e\Vds = 1/16$. With increasing $\lambda$, the \emph{less active} rare trajectories show a continuous transition boundary terminating in a critical point close to $s=0$ while the \emph{more active} dynamics exhibit a series of first-order transitions which become diffuse as $s$ is increased to zero.  Cuts through the plot show the activity as a function of $\lambda$ for (b) $s=-0.1$, (c) $s=0$ and (d) $s=0.1$.  Samples of the $s=0$ quantum-jump trajectories for quanta leaving the resonator are shown for couplings (e) $\lambda=0.056$, (f) $\lambda=0.041$ and for (g) $\lambda=0.020$ where the system is close to the critical point shown in (a).}
   \label{EJ1_kappa}
 \end{centering}
\end{figure}

For ensembles of trajectories with $s\ne 0$, we will use the activity at non-zero $s$, found from $k(s) = -\theta'(s)$ by extending~\eqref{eq:act} to non-zero $s$, as the order parameter to distinguish dynamical phases.  In Fig.~\ref{EJ1_kappa}, $k(s)$ is used to construct the $\lambda-s$ phase diagram, shown for SSET parameters~\cite{Blencowe2005} $\Delta E / eV_{\text{ds}} = -0.1$ and $E_J/eV_{\text{ds}} = 1/16$.
Figure~\ref{EJ1_kappa} demonstrates how the signatures of dynamical phase transitions in the $s=0$ physical dynamics shown in Fig.~\ref{EJ1_s0} may be related to the full $\lambda-s$ phase diagram.  While at $s=0$ we see smooth crossovers in the dynamical behavior, for ensembles biased towards rare trajectories the crossovers become sharp. The less active trajectories for $s<0$ show a series of first-order dynamical transitions where the activity shows distinct jumps at singular points in the $\lambda-s$ plane.  For $s>0$, the more active trajectories undergo a single phase transition provided that $s>s_c$, where $s_c>0$ is the $s$-coordinate of a critical point at which the right-most transition line terminates when $\lambda \simeq 0.02$.  We illustrate the connection between the dynamical phase diagram and the $s=0$ trajectories by plotting examples of trajectories at different $\lambda$ in Fig.~\ref{EJ1_kappa}.  These were found by sampling the full distribution of counting events in the non-equilibrium steady state.  In particular, we note the development of fluctuations in the activity on a wide range of timescales close to the critical point near $\lambda=0.02$. 

The similarity of the dynamical phase diagram in Fig.~\ref{EJ1_kappa} to that derived for the micromaser (Fig. 1 in ~\cite{Garrahan2011}) further demonstrates the parallels between the SSET-resonator system and the micromaser noted in~\cite{Rodrigues2007}.  However, the SSET-resonator system allows a broader range of dynamical behavior which can be found by varying its large number of tunable parameters.  In particular, we can vary the strength of coherent oscillations in the SSET by altering the Josephson energy $E_J$, and study the corresponding dynamical states of the resonator.

\begin{figure}[tb]
 \begin{centering}
   \includegraphics[scale=1.2]{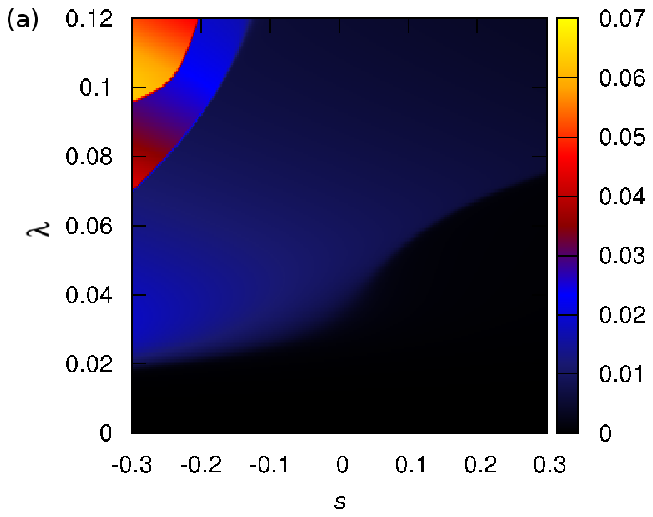}
   \includegraphics[scale=1.2]{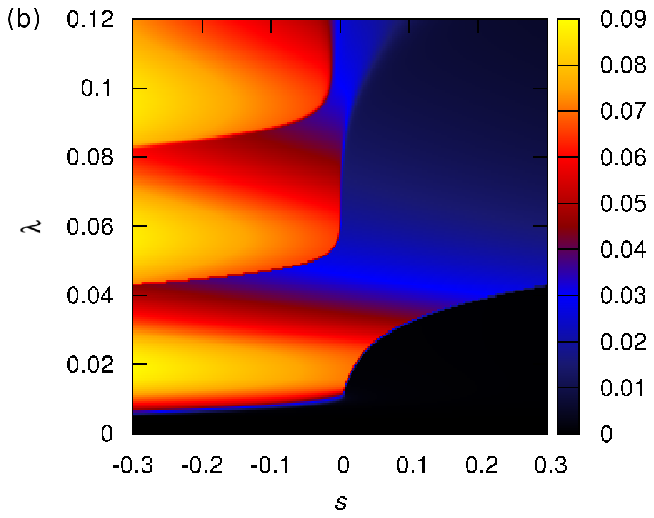}
   \caption{(color online) Plots of the activity $k$ as a function of coupling strength $\lambda$ and counting field $s$ for system parameters in Fig.~\ref{EJ1_kappa} but with (a) $E_J/eV_\text{ds} = 1/32$ and (b) $E_J/eV_\text{ds}=1/8$ for fixed $\Gamma = eV_\text{ds}/h$.}
   \label{EJ2_kappa}
 \end{centering}
\end{figure}

In Fig.~\ref{EJ2_kappa}, phase diagrams equivalent to Fig.~\ref{EJ1_kappa} are shown for Josephson energies both increased and decreased by a factor of two.
For the case of the smaller Josephson energy $E_J/eV_\text{ds} = 1/32$, the $s=0$ dynamics appears uninteresting.  As $\lambda$ is increased from zero, the activity shows a weaker version of the continuous crossover seen for smaller $\lambda$ in Fig.~\ref{EJ1_kappa}a, but none of the first-order crossovers.  However, the other dynamical phases which occur when the resonator is oscillating in a limit cycle may still be seen when biasing the system towards more active rare trajectories.  We can interpret these findings from the $s$-ensemble for the physical dynamics as follows: while on average the smaller Josephson energy restricts the ability of the SSET to populate the oscillator with a large number of quanta, on rare occasions the resonator will be excited to one of the limit cycle states where it will release photons into its environment more often.  So when biasing the ensemble of trajectories increasingly towards those with higher activity, first-order phase boundaries are crossed and we may infer that the typical state of the resonator for this biased ensemble must be higher in energy than for the $s=0$ case of the same $\lambda$.  We note that the biased ensembles of trajectories, where $s\ne 0$, can be inferred from accurate measurements of the distribution of trajectories at $s=0$: with a knowledge of the large-deviation function $\varphi$ in~\eqref{eq:trajdist}, $\theta(s)$ can be derived by Legendre transform, with $k(s)$ then found by differentiation.  In this indirect way, the full phase diagram could in principle be generated from measurements of the physical system.  We are therefore able to give a thermodynamic interpretation of the distribution of trajectories which gives insights into the underlying internal dynamics.  
However, since the biased ensembles may be generated from a knowledge of the full distribution of trajectories at $s=0$, the fluctuations at $s=0$ must reflect the structure of the phase diagram found when favoring rare trajectories.

We now study the case where $E_J$ is increased from the micromaser-like regime illustrated in Fig.~\ref{EJ1_kappa}.  We show that this introduces new features in the dynamical phase diagram even at $s=0$.  For the Josephson energy $E_J/eV_\text{ds}=1/8$ shown in Fig.~\ref{EJ2_kappa}, the dynamical transitions remain sharp close to $s=0$.  This is in contrast to Fig.~\ref{EJ1_kappa} where the transition line where $s<0$ gradually becomes diffuse as it approaches $s=0$.  In Fig.~\ref{EJ2_kappa}, there is also a new crossover in dynamical behavior for $s>0$ at coupling strengths larger than for the initial continuous transition.  The first-order transition line which approaches from $s<0$ extends to $s=0$, but becomes diffuse for positive values of $s$.  These interesting new features go beyond what can be seen in the micromaser~\cite{Garrahan2011} and their origin will be explained using a mean-field treatment in Section~\ref{Mean}.  In general, we observe that the sharpness of the transitions near $s=0$ increases with the Josephson energy.  Correspondingly, the number of quanta in the resonator state increases, such that these increases in $E_J$ take the system closer to the thermodynamic limit for dynamical phase transitions.

\section{Mean-Field Approach}
\label{Mean}

In this section we will first present an overview of mean-field results~\cite{Rodrigues2007a,Marquardt2006},  as applied to the SSET-resonator system with the Lindblad baths used in this paper.  These results provide a mean-field theory for the resonator amplitude at $s=0$ in terms of an effective damping, non-linear in the resonator amplitude, arising from coupling to the SSET.  From this we will develop an effective stochastic Liouvillian superoperator for the resonator, from which the behavior at $s\ne 0 $ will be derived.  We will then relate these mean-field solutions to the exact results found from numerical diagonalization of the full generalized master equation~\eqref{eq:Ws_osc} and explore the origin of the dynamical phase boundaries in terms of multistabilities in the resonator.

\subsection{Mean-field Dynamics at $s=0$}

At $s=0$, the dynamical instabilities of the SSET-resonator system may be understood from a mean-field description of the dynamics.  We follow the approach developed in Ref.~\cite{Rodrigues2007a}. Although the form of the bath for the resonator we use differs from that in~\cite{Rodrigues2007a}, the final mean-field descriptions are the same.

  We define the occupation probabilities of the SSET states $\ket{0}$, $\ket{1}$ and $\ket{2}$ respectively by $p_{00}$, $p_{11}$ and $p_{22}$ and the off-diagonal element of the SSET density matrix describing coherences between states $\ket{0}$ and $\ket{2}$ by $p_{02}$.  The mean-field equations of motion for the SSET are given by
\begin{eqnarray}
\expectation{\dot{p}_{11}} &=& \Gamma\left(\expectation{p_{22}} - \expectation{p_{11}}\right) \label{eq:mf1}\\
\expectation{\dot{p}_{22}} &=& -\Gamma\expectation{p_{22}} + \frac{i E_J}{2 \hbar}(\expectation{p_{02}} - \expectation{p_{20}}) \label{eq:mf1b}\\
\expectation{\dot{p}_{02}} &=& \frac{iE_J}{2 \hbar}(2\expectation{p_{22}} + \expectation{p_{11}} -1) - \frac{\Gamma}{2}\expectation{p_{02}} \nonumber \\
&+& \frac{i}{\hbar}\left[\Delta E + 2x_s  C  \left(\expectation{a}+\expectation{a^\dagger}\right)  \right]  \expectation{p_{02}} \quad \label{eq:p02}
\end{eqnarray}
with the mean-field resonator dynamics described by
\begin{eqnarray}
\!\!\expectation{\dot{a}} \! &=&\! -i\omega \expectation{a} - i x_s  C(\expectation{p_{11}}\!\!+\!2\expectation{p_{22}})  -\frac{\gammaext}{2}\!\expectation{a}  \label{eq:mf2a}\\
\!\!\expectation{\dot{a}^\dagger} \! &=&\!  i\omega \expectation{a^\dagger}\! +\! i x_s  C(\expectation{p_{11}}+2\expectation{p_{22}}) - \frac{\gammaext
}{2}\!\expectation{a^\dagger}. \label{eq:mf2}
\end{eqnarray}
The occupation of the SSET state $\ket{0}$ is determined by the other two SSET states since $\expectation{p_{00}}+\expectation{p_{11}}+\expectation{p_{22}}=1$.  We have replaced correlators $\expectation{p_{02}\, a^{(\dagger)}}$ with $\expectation{p_{02}}\expectation{a^{(\dagger)}}$ in~\eqref{eq:p02}.  This approximation does not fully capture the quantum dynamics, but comparisons with numerics~\cite{Rodrigues2007a} show that it provides a reasonably accurate picture of the average dynamics of the system, including the limit-cycle states of the resonator.

Because the SSET-resonator coupling is weak, we expect the resonator to undergo harmonic oscillations at its fundamental frequency $\omega$ with an amplitude $A$ which changes on time scales much larger than $\omega^{-1}$.  We will denote the position about which the resonator oscillates (often described as its average or fixed-point position) by $\xfp$, which itself is a function of the SSET parameters and the coupling strength.  We will solve for the SSET dynamics assuming such harmonic oscillations of arbitrary amplitude in the resonator and then use these results to find the stable non-equilibrium steady states of the resonator.  We therefore proceed to solve for the SSET dynamics by choosing the following ansatz for the resonator state:
\begin{eqnarray}
\expectation{a} &=& \sqrt{\frac{m\omega}{2\hbar}}(\xfp + A e^{\omega t})\nonumber  \\
\expectation{a^\dagger} &=& \sqrt{\frac{m\omega}{2\hbar}}(\xfp + A e^{-\omega t})\,. \label{eq:ansatz}
\end{eqnarray}
Because $E_J/\hbar\Gamma\ll 1$, we may also consider that the occupation probabilities $\expectation{p_{11}}$ and $\expectation{p_{22}}$ will remain much less than unity~\cite{Rodrigues2007a} and can be neglected in~\eqref{eq:p02}.  When using the ansatz~\eqref{eq:ansatz}, Eq.~\eqref{eq:p02} becomes
\begin{eqnarray}
\expectation{\dot{p}_{02}} &=& \left\{i\left[\frac{\Delta E}{\hbar} + \frac{2m\omega^2 x_s}{\hbar}(\xfp + A\cos\phi)\right]-\frac{\Gamma}{2}\right\}\expectation{p_{02}} \nonumber \\
 &-& \frac{iE_J}{2\hbar}\,.
 \label{eq:p022}
\end{eqnarray}
which can be solved in closed form for a given resonator amplitude $A$.  
We absorb the $A$-dependence of $\expectation{p_{02}}$ in a time-dependent phase shift $\alpha(t) = (2 m \omega A x_s/\hbar) \sin \omega t$ and then find the Fourier series such that $\expectation{p_{02}} = e^{i\alpha(t)}\sum_n e^{i\omega n t} \tilde{p}^n_{02}$~\cite{Marquardt2006}.
We find Fourier coefficients $\tilde{p}^n_{02} = \psi^n J_n(-z)$, where $J_n(z)$ are Bessel functions of the first kind, and the parameters $\psi^n$ are given by
\begin{equation}
\psi^n = \frac{-iE_J/\hbar}{2[i\omega n -i(\Delta E/\hbar + 2m\omega^2 x_s \xfp/\hbar-\Gamma/2)]}\,.
\label{eq:psi}
\end{equation}

We now turn back to the resonator dynamics.  With our ansatz~\eqref{eq:ansatz}, we use Eqs.~\eqref{eq:mf2a} and~\eqref{eq:mf2} to find an equation of motion for the resonator amplitude $A$.  Since the amplitude changes on timescales much longer than $\omega^{-1}$, we average $A$ over a time period $2\pi/\omega$ to find
\begin{equation}
\dot{\bar{A}} = -\frac{\gammaext}{2}\bar{A} -\omega x_s \frac{1}{2\pi} \int_{0}^{2\pi/\omega}\!\! dt (\expectation{p_{11}} + 2\expectation{p_{22}}) \sin \omega t
\label{eq:Adot2}
\end{equation}
where the bar on the amplitude signifies that the relation only holds on time scales long compared with $\omega^{-1}$.  The second term on the right-hand side may be considered an effective, amplitude-dependent damping term $\gamma_\text{SSET}(A) A$, but it should be noted that when $\Delta E<0$, $\gamma_\text{SSET}$ is actually negative as the SSET pumps the resonator.  Using Fourier series of the form $\expectation{p_{11}} = \sum_n e^{i\omega n t} p^n_{11}$ and $\expectation{p_{22}} = \sum_n e^{i\omega n t} p^n_{22}$ we find that 
\begin{equation}
\gammaSSET(A) A = -\frac{i \omega x_s}{2} \left((p_{11}^1 - p_{11}^{-1}) + 2 (p_{22}^{1} - p_{22}^{-1})\right)\,.
\end{equation}
This may be found using the Fourier transforms of Eqs.~\eqref{eq:mf1} and~\eqref{eq:mf1b} and Eq.~\eqref{eq:p022}.  Ultimately we find
\begin{equation}
\gamma_\text{SSET}A = -\frac{\omega x_s E_J}{\hbar} \text{Im}\left[\left(\frac{2}{\Gamma+i\omega} + \frac{\Gamma}{(\Gamma+i\omega)^2}\right)\beta\right]\, ,
\label{eq:gSSET}
\end{equation}
where
\begin{equation}
\beta = \frac{1}{2i}\sum_m\left(\psi^{-m}J_{m+1}(z) - (\psi^{-m})^* J_{m-1}(z)\right)J_m(z)\,.
\label{eq:beta}
\end{equation}
This complicated form of $\beta$ arises from the need to switch between the phase-shifted Fourier series for $\expectation{p_{02}}$ and the Fourier series without phase shifts.  The $s=0$ steady states are found as solutions where $\dot{\bar{A}}=0$.  

As well as finding the steady-state amplitudes of the resonator, we may also find $\xfp$.  While the effective damping is derived from the lowest frequency oscillating terms in the Fourier series for the charge dynamics, the fixed-point position may be found in terms of the zero-frequency terms.  Using again the Fourier series for Eqs.~\eqref{eq:mf1} and~\eqref{eq:mf1b} and Eq.~\eqref{eq:p022}, we find 
\begin{equation}
\xfp = -x_s(p_{11}^0 + 2p_{22}^0) = -\frac{3E_J^2 \,x_s}{2\Gamma\, \Delta E^2 + 3E_J^2 + \Gamma^2}\,.
\label{eq:xfp}
\end{equation}

In order to find mean-field solutions when $s\ne 0$, we would need to repeat this analysis with $s$ introduced from the start.  However, since the full $s$-dependent master equation~\eqref{eq:W} is not trace preserving for general $s$, the method above is not easy to extend to non-zero $s$.  Instead we proceed via a different route: we construct an effective master equation for the system with a simpler structure.  This new master equation is chosen to give the mean-field results at $s=0$ found from the solution above and allows solutions with $s\ne 0$ to be found using a variational method.

\subsection{Effective Stochastic Master Equation}

Using the above result for the effective (negative) damping term for the resonator, arising from the weak coupling to the SSET island, we now construct an effective master equation for the resonator.  By construction, this master equation will correctly describe the average $s=0$ dynamics to within the mean-field approximations above.  However, noise arising from the stochastic quasiparticle decay process cannot be captured.  The structure of the master equation is simpler as it just involves two Lindblad operators;  one is associated with the environment of the resonator and the other describes the driving due to the SSET.  The complexity of the mean field solution will be contained in these driving terms, which must be amplitude dependent.   By applying the counting field $s$ to the trajectories defined by the time record of quanta entering the environment from the oscillator, this generalized master equation is
\begin{eqnarray}
\dot{\rho_s}=\tilde{\mathcal{W}}_s(\rho_s) &=& L\rho_s L^\dagger -\frac{1}{2}\{L^\dagger L,\rho_s\}\nonumber\\
 &+& \gammaext \left(e^{-s} a\rho_s a^\dagger - \frac{1}{2}\{a^\dagger a,\rho_s\} \right)
 \label{eq:effWs}
\end{eqnarray}
where the new Lindblad operator is defined by
\begin{equation}
L = \sqrt{g(n)}\, a^\dagger
\end{equation}
where $n=a^\dagger a$ and
\begin{equation}
g(n) = -\gamma_\text{SSET}(n) \,.
\end{equation}
The amplitude-dependent negative damping allows construction of an $n$-dependent driving term, where $n$ is the number of quanta in the oscillator state, which is related to its amplitude via $n=A^2(m\omega/2\hbar)$.  In addition to its dependence on $n$, $g(n)$ depends implicitly on the SSET parameters and the coupling strength, $\lambda$.  Because this stochastic description only couples diagonal elements of $\rho_s$ to other diagonal elements in the basis of number states for the harmonic oscillator, we may construct from~\eqref{eq:effWs} a normal operator $W_s$.  This operator acts on diagonal density matrices in the number basis, which we may represent as vectors whose time evolution follows:
\begin{eqnarray}
\dot{\rho}_s &=& W_s{\rho}_s = \Big\{e^{-s}\gammaext \sqrt{{a}^\dagger {a}+1} \, {a} -\gammaext {a}^\dagger {a} \nonumber \\
&+& g(n)\sqrt{{a}^\dagger {a}+1}\, {a}^\dagger - g(n)({a}^\dagger {a}+1)\Big\}{\rho}_s\,.
\end{eqnarray}
 Using a variational ansatz of coherent states as in~\cite{Garrahan2011}, we set ${a} = e^{i\delta} \sqrt{n}$ and ${a}^\dagger = e^{-i\delta}\sqrt{n}$.  We can then find a mean-field estimate of the largest eigenvalue, $\theta(s)$, by maximizing $W_s$ with respect to $\delta$ and $n$.  This is done by finding the values of $\delta$ and $n$ which satisfy $\partial W_s/\partial \delta =0$ and $\partial W_s/\partial n =0$.  The first equation allows $\delta$ to be eliminated so that
\begin{equation}
{a} = e^{-s/2} \sqrt{\frac{n\, g(n)}{\gammaext}}
\end{equation}
and ${a}^\dagger = n/{a}$.  Inserting these into $W_s$ yields
\begin{equation}
W_s = 2e^{-s/2} \sqrt{n(n+1)\gammaext g(n)} - \gammaext n - g(n)(n+1)
\label{eq:Ws_f}
\end{equation}
which may be maximized with respect to $n$ numerically.  

\begin{figure}[b]
 \begin{centering}
   \includegraphics[scale=1.2]{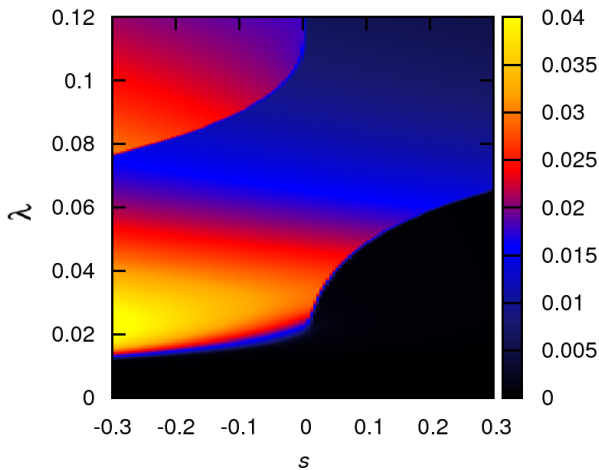}
   \caption{(color online) Mean field: A plot of the activity $k$ as a function of coupling strength $\lambda$ and counting field $s$ for system parameters $\gammaext/\Gamma = 0.0005$, $\omega/\Gamma = 1$, $\Delta E / eV_\text{ds} = -0.1$, $E_J/eV_\text{ds} = 1/16$ for fixed $\Gamma = eV_\text{ds}/h$.}
   \label{EJ1_mf}
 \end{centering}
\end{figure}

\begin{figure}[htb]
 \begin{centering}
   \includegraphics[scale=1.2]{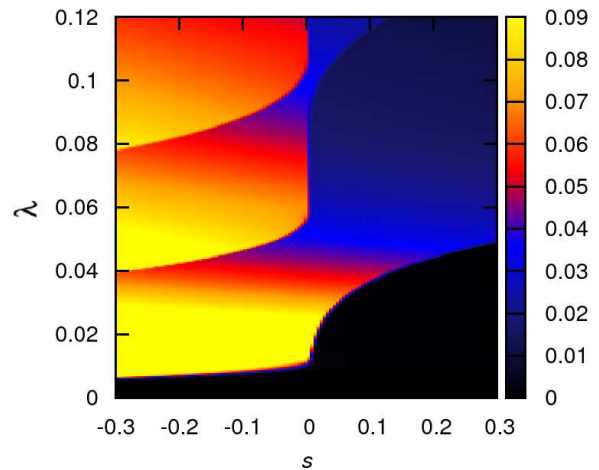}
   \caption{(color online) Mean field: A plot of the activity $k$ as a function of coupling strength $\lambda$ and counting field $s$ for system parameters $\gammaext/\Gamma = 0.0005$, $\omega/\Gamma = 1$, $\Delta E / eV_\text{ds} = -0.1$, $E_J/eV_\text{ds} = 1/8$ for fixed $\Gamma = eV_\text{ds}/h$.}
   \label{EJ2_mf}
 \end{centering}
\end{figure}

For the SSET parameters which show the micromaser-like behavior demonstrated in Fig.~\ref{EJ1_kappa}, the mean-field solution found from Eq.~\eqref{eq:Ws_f} is shown in Fig.~\ref{EJ1_mf}.   The case of larger $E_J/eV_\text{ds}=1/8$ shown in Fig.~\ref{EJ2_kappa} may be compared with the mean-field results in Fig.~\ref{EJ2_mf}.
For larger $E_J$ there is excellent agreement, but for lower $E_J$ values the phase diagrams, though qualitatively similar, differ quite noticeably in the exact locations of transitions.  (Accordingly, the mean-field result captures the dynamical phase diagram less accurately when $E_J/e\Vds = 1/32$.) 
We suggest that this is most likely due to the SSET charge fluctuations which are neglected in Eq.~\eqref{eq:effWs}.  In the parameter regime studied here, these fluctuations are more significant when $E_J$ is smaller.
Support for this view comes from the numerical result for the full distribution $P(n)$ for the occupation of the resonator number states for $E_J/e\Vds = 1/8$ shown in Fig.~\ref{EJ2_s}(b).  This should be contrasted with the $P(n)$ distribution in Fig.~\ref{EJ1_s0} for $E_J/e\Vds = 1/16$, which shows significant blurring between the peaks in $n$ corresponding to different limit cycles, when compared to Fig.~\ref{EJ2_s}(b).  These effects will not be captured by the mean-field solution as it neglects these fluctuations.

Another observation we make is that the mean-field phase diagrams show an extra phase boundary when $s>0$ for $E_J/e\Vds = 1/8$, which is also present in the numerical result in Fig.~\ref{EJ2_kappa}.  We will show that the appearance of this new phase boundary may be understood in terms of the mean-field equations involving the non-linear driving term due to the SSET. 

\subsection{Phase Transitions at $s=0$}

At $s=0$, the dynamical states of the resonator are a set of limit cycles whose amplitudes are determined by balancing the effects of the dissipation due to the environment, which is linear in the resonator amplitude, and the non-linear driving due to the SSET.  This is clear from Eq.~\eqref{eq:Ws_f} when written in the form
\begin{eqnarray}
W_s = &-&(\sqrt{\gammaext n} - \sqrt{g(n) (n+1)})^2 \nonumber \\
&&+ 2(e^{-s/2}-1)\sqrt{n(n+1)\gammaext g(n)}\,.
\label{eq:Ws_f2}
\end{eqnarray}
At $s=0$ we require $\theta(0)=0$.  This occurs when Eq.~\eqref{eq:Ws_f2} is maximized since, for $s=0$, $W_s\le 0$.  Multistabilities for the resonator state occur when there is more than one value of $n$ which satisfies $W_s = 0$.  From Eq.~\eqref{eq:Ws_f2}, we can also see that these multistabilities are associated with phase boundaries at $s=0$.  To see this, consider the case where many values $n_i$, with $n_{i+1}>n_i$, satisfy $W_{s}=0$ with $s=0$.  Now, noting that $g(n)$ is a smooth function of $n$, if we perturb $s$ from zero by a small amount $\delta s$, changes in the positions of maxima $n_i$ will be of order $\delta s$ and, therefore, much smaller than $n_{i+1}-n_i$.  Now if we consider the values of $W_s$ when introducing small shifts $\delta s$ at the maxima $n_i$, we see that 
\begin{equation}
W_{\delta s} = -\delta s \sqrt{\gammaext g(n_i) n_i (n_i+1)} = \delta s \,\gammaext\, n_i
\label{eq:deltas}
\end{equation}
to first order in $\delta s$.  From Eq.~\eqref{eq:deltas} we can see that if $\delta s>0$, maximization of $W_s$ is found by selecting the smallest $n_i$.  Conversely, for $\delta s<0$, $W_s$ is maximized by picking the largest $n_i$.   Therefore we understand multistabilities, where there are two or more solutions to the $s=0$ mean-field equations, are associated with $s=0$ mean-field dynamical phase boundaries.  Upon changing $\lambda$, phase boundaries extend into the $s>0$ or $s<0$ half-planes when, respectively, the smallest  $n_i$ solution or the largest $n_i$ solution changes.  With this insight, we now turn to the question of why increasing $E_J$ introduces a new phase boundary on the less active side of the phase diagram.

\begin{figure}[htb]
 \begin{centering}
   \includegraphics[scale=1.3]{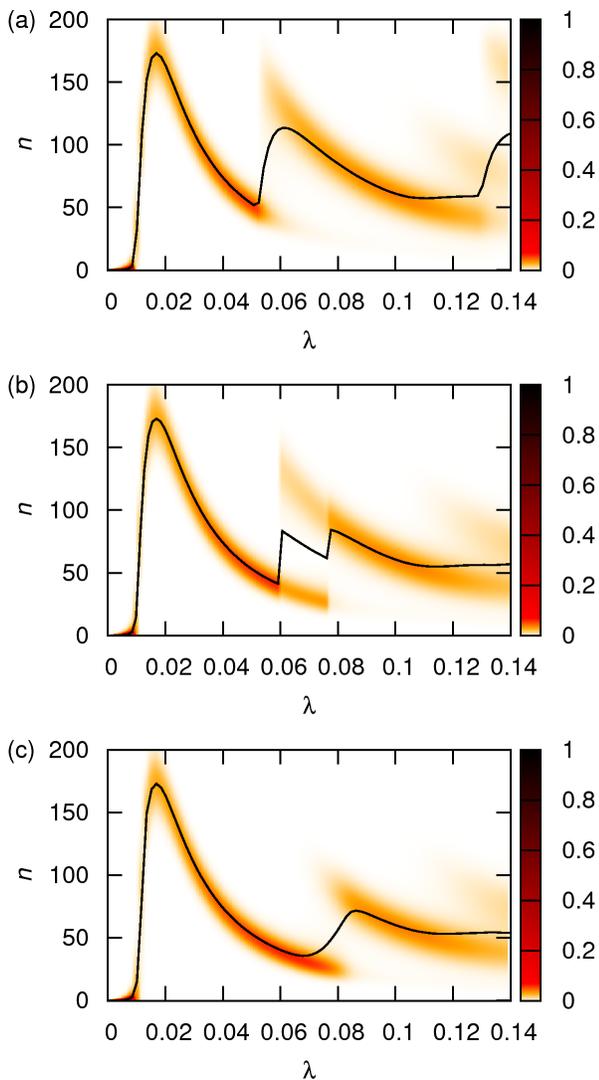}
   \caption{(color online) The state of the resonator close to $s=0$. Full distributions of $P(n)$ are plotted as a function of $\lambda$ for the parameters in Fig.~\ref{EJ2_kappa} with $E_J/V_ds = 1/8$.  Shown are three $s$ values: (a) $s=-0.005$, (b) $s=0$ and  (c) $s=0.005$.  Shown with solid lines are the expectation values $\expectation{n}$.  }
   \label{EJ2_s}
 \end{centering}
\end{figure}

For $E_J/e\Vds = 1/16$, as shown in Fig.~\ref{EJ1_s0}, the onset of bistability occurs at $\lambda \simeq 0.012$ and the resonator remains multistable as $\lambda$ is increased further.   In contrast, when $E_J/e\Vds = 1/8$, there exists a second region of monostablilty as $\lambda$ is increased beyond the bistable region.  In Fig.~\ref{EJ2_s}, we show the full $P(n)$ distribution for the occupation of the oscillator number states when $E_J/e\Vds = 1/8$, obtained from exact diagonalization of the generalized master operator.  The initial onset of bistability occurs at $\lambda \simeq 0.06$ with monostability again beyond $\lambda \simeq 0.077$.  This is consistent with the phase boundary shown at $s=0$ between these values of $\lambda$ in both the exact phase diagram in Fig.~\ref{EJ2_kappa} and the mean-field phase diagram in Fig.~\ref{EJ2_mf}.  In Fig.~\ref{EJ2_s}, we also show that for $s$ perturbed slightly from the zero, the bistability disappears from the full $P(n)$ distribution as one of the two possible limit cycles is selected, consistent with Eq.~\eqref{eq:Ws_f2}.

\begin{figure}[htb]
 \begin{centering}
   \includegraphics[scale=1.4]{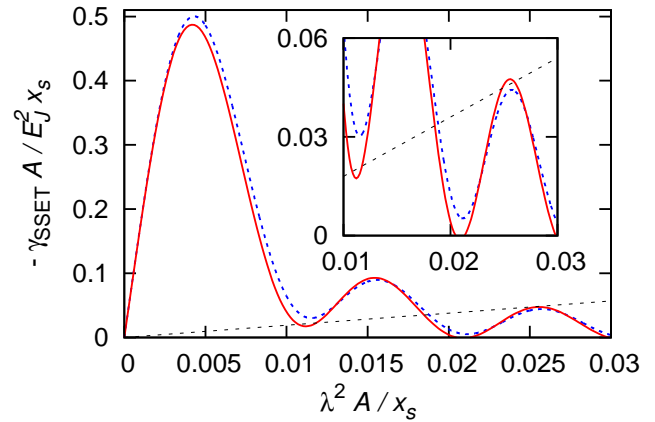}
   \caption{Plots of the effective negative damping on the resonator due to the charge on the SSET island.  Plotted are the damping curves corresponding to Josephson energies $E_J/e\Vds = 1/16$ (solid) and $E_J/e\Vds = 1/8$ (dashed), with a magnified plot of the area of interest shown inset.  All other parameters are as in Figs.~\ref{EJ1_mf} and~\ref{EJ2_mf}.  The dotted line of gradient $\gammaext/ E_J^2 \lambda^2 = 1.8$ illustrates the origin of the second region of monostability when $E_J/e\Vds = 1/8$ since this line has just one intersection with the dashed curve.}
   \label{gSSET}
 \end{centering}
\end{figure}

We now turn specifically to the origin of the second monostability present for $E_J/e\Vds = 1/8$.  We may understand this with reference to the effective negative damping $\gammaSSET$ on the oscillator provided by the SSET.  If we neglect the effects of the SSET on the fixed-point displacement of the resonator, $\xfp$, from Eqs.~\eqref{eq:psi},~\eqref{eq:gSSET} and~\eqref{eq:beta}, we note that $\gammaSSET A/x_s$ scales with $E_J^2$ so that $\gammaSSET A/ E_J^2 x_s$ is a function of $\lambda A/x_s$.  Therefore, where $\xfp$ may be neglected, we expect the same structure of limit cycles and multistabilities at different $E_J$ from the mean-field theory.  However, if we include the effects of the modified fixed point using Eq.~\eqref{eq:xfp}, we find that the effective damping is modified by a small but significant amount.  In Fig.~\ref{gSSET}, we show how $\xfp$ perturbs the form of $\gammaSSET$ upon increasing $E_J$.  By plotting $-\gammaSSET A/ E_J^2 x_s$ against $\lambda A/x_s$, we show disappearance of bistability when increasing $E_J/e\Vds$ from $1/16$ to $1/8$.  Limit cycle solutions occur where $-\gammaSSET = \gammaext$.  These solutions are illustrated on Fig.~\ref{gSSET} by the intersection of $-\gammaSSET A/ E_J^2 x_s$ with lines of constant gradient $\gammaext/ E_J^2 \lambda^2$.  To understand the existence of new limit cycles upon increasing $\lambda$, it should be noted that these occur when $\lambda$ is large enough that such lines have shallow enough slope to intersect peaks in the effective damping at larger $\lambda^2 A/x_s$.  The case where $\gammaext/ E_J^2 \lambda^2 = 1.8$ is plotted to illustrate that while monostability exists beyond the first region of bistability when $E_J/e\Vds = 1/8$, for $E_J/e\Vds = 1/16$ these solutions are not possible due to the smaller ratio $\xfp/x_s$.

\section{Quasiparticle Trajectories}
\label{Quasi}

\begin{figure}[htb]
 \begin{centering}
   \includegraphics[scale=1.2]{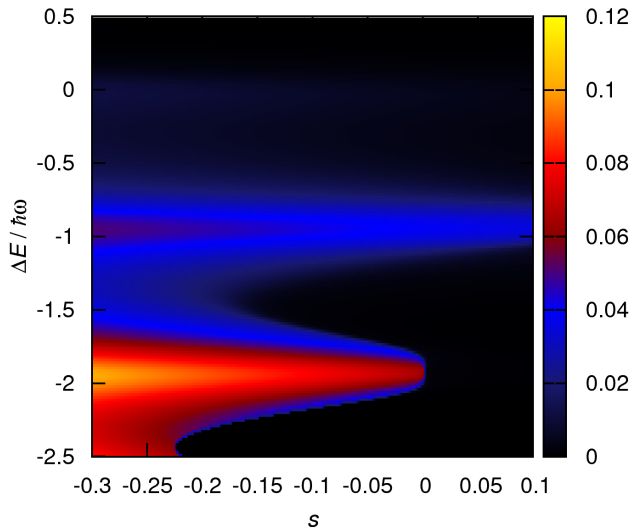}
   \caption{(color online) A plot of the $\Delta E -s$ phase diagram showing the activity for quantum jumps associated with the environment of the resonator.  The vertical axis shows $\Delta E /\hbar\omega$.  The system parameters are $\omega/\Gamma = 3.5$, $\gammaext/\Gamma = 0.002$, $\lambda = 0.2$ with Josephson energy $E_J/\Vds = 1/16$.}
   \label{DeltaE}
 \end{centering}
\end{figure}

We now turn to the trajectories of quasiparticles decaying into the right-hand lead of the SSET.  For oscillators with $\omega/\Gamma \le 1$, there are no features in the current corresponding to changes in the dynamical state of the resonator. 
So far we have considered the case of a resonator with relative frequency $\omega/\Gamma =1$.  In this regime, we are not able to resolve individual features in the resonator energy as a function of $\Delta E$, which arise from inelastic absorption of energy quanta from the SSET when $\Delta E = -j\hbar \omega$, for integral $j>0$.  However, if we consider a fast oscillator where $\omega/\Gamma \gg 1$, the timescales for the SSET and resonator dynamics become separated and the oscillator does show distinct peaks in energy when $\Delta E = -j\hbar \omega$~\cite{Rodrigues2007}.  Between these peaks, the SSET has little influence on the resonator dynamics.  In Fig.~\ref{DeltaE}, we plot numerical results for the $\Delta E-s$ phase diagram in this regime.  The resonator absorption peaks separated by $\hbar \omega$ correspond to strong features in the phase diagram.  We have also applied the mean field theory developed in Section~\ref{Mean} to this parameter regime.  This method also exhibits features separated by $\hbar \omega$ but only shows qualitative agreement with Fig.~\ref{DeltaE}, a result which we attribute to the small resonator energy across much of the $\Delta E - s$ phase diagram in this case.

The quasiparticle current in the resolved side-band limit, occurring when $\omega/\Gamma \gg 1$, does exhibit maxima which coincide with peaks in the energy of the resonator as a function of $\Delta E$~\cite{Harvey2008}.  We understand the emergence of these peaks as follows:  when $\omega/\Gamma \gg 1$,  the separation of timescales of resonator and SSET means that for most bias points, the quasiparticle current is low and not strongly influenced by the presence of the resonator.  However, at the particular values $\Delta E = -j\hbar\omega$, absorption of energy by the resonator from the charges occurs allowing a greatly enhanced charge flow through the SSET island (analogous to an inelastic tunneling current).  This correlation between resonator energy and charge dynamics motivates our study of the trajectories of SSET quasiparticles.

We now apply the $s$-ensemble to charges leaving the SSET island.  We wish to explore the extent to which the trajectories of the SSET quasiparticles mirror the trajectories of the photons entering the environment of the resonator.  To do this, we apply the $s$-ensemble to the counting of charges leaving the SSET island by diagonalizing the superoperator $\mathcal{W}_s^\text{q.p.}(\rho)$ defined by
\begin{equation}
\mathcal{W}_s^{\text{q.p.}}(\rho) = -\frac{i}{\hbar} [H,\rho] + \mathcal{L}_s^\text{q.p.}(\rho) + \mathcal{L}^\text{osc.}(\rho)
\label{eq:Ws_qp}
\end{equation}
where
\begin{eqnarray}
\mathcal{L}^{\text{q.p.}}_s(\rho) = &\Gamma e^{-s} (\ket{1}\bra{2}\rho\ket{2}\bra{1} + \ket{0}\bra{1}\rho\ket{1}\bra{0}) \nonumber \\
&- \frac{\Gamma}{2}\left(\{\ket{2}\bra{2},\rho \} + \{\ket{1}\bra{1},\rho \}\right)\,.
\end{eqnarray}
The large-deviation function for quasiparticle trajectories, $\theta_\text{q.p.}(s)$, is given by the largest real eigenvalue of $\mathcal{W}_s^\text{q.p.}(\rho)$.  We will study the dynamics of SSET charges using the quasiparticle activity $-\partial\theta_\text{q.p.}(s)/\partial s$.  

\begin{figure}[htb]
 \begin{centering}
   \includegraphics[scale=1.2]{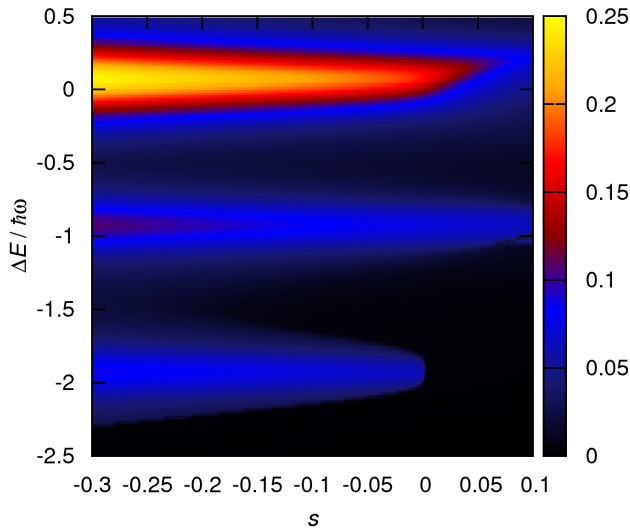}
   \caption{(color online) A plot of the $\Delta E -s$ phase diagram using the quasiparticle activity, found from the generalised master equation~\eqref{eq:Ws_qp}.  The vertical axis shows $\Delta E /\hbar\omega$.  The system parameters are identical to those in Fig.~\ref{DeltaE}.}
   \label{DeltaESSET}
 \end{centering}
\end{figure}

For a resonator where $\omega/\Gamma \apprle 1$, we see no transitions in the dynamical behavior of the SSET charges, consistent with the known $s=0$ behavior~\cite{Harvey2008}. However, for the system parameters in Fig.~\ref{DeltaE} corresponding to a fast oscillator, we find that the $\Delta E-s$ phase diagram, shown in Fig.~\ref{DeltaESSET}, shows similar structure to the dynamical phase diagram for the trajectories of quanta leaving the resonator in Fig.~\ref{DeltaE}.    In the resolved side bands, where multiples of the oscillator level spacing closely match the level spacing of the SSET charge states, the inelastic interaction between SSET and resonator enhances the rate at which quasiparticles decay.  Therefore the trajectories of both resonator photons and charge quasiparticles allow inference of the resonator state in this regime.  However, the peaks in the respective activities in Figs.~\ref{DeltaE} and~\ref{DeltaESSET} at $\Delta E = 0$ show a marked difference.  When $\Delta E=0$, there is no significant driving of the resonator by the SSET and so the number of quanta leaving the resonator is small.  In contrast, the decay of quasiparticles into the right-hand lead is large, such that the activity when counting charge quanta is large at $\Delta E =0$.  Finally, we mention another interesting feature of Fig.~\ref{DeltaE}.  When biasing towards more-active trajectories, we find that the activity associated with counting quanta from the resonator is bigger when $\Delta E = -2\hbar\omega$ than when $\Delta E =-\hbar\omega$.  This effect is only seen for biased ensembles with $s<0$.

\section{Conclusions}
\label{Conclusions}

In this paper we have used the $s$-ensemble approach to study the dynamics of a complex system with two components, each of which has non-trivial dynamics.  The coupling of an SSET and a resonator has been studied using mean-field techniques previously~\cite{Rodrigues2007a}.  We devised a method to extend a standard mean-field theory for the coupled system to enable us to study ensembles of trajectories with non-zero $s$.  Comparing the results of our mean-field theory to those employing essentially exact numerical diagonalization of the generalized master equation, we found good agreement for system parameters where the resonator is driven into high energy states of self-sustaining oscillation. 

We find the accuracy of mean-field results improves for large Josephson energy, where the resonator energy is large.  We were able to use our method to understand the correspondence between multistabilities in the state of the resonator and boundaries in the dynamical phase diagram at $s=0$.  This demonstrates how the dynamical phase diagram for trajectories can be used to infer properties of the internal dynamics of the system.  However, even when there are no phase boundaries near $s=0$, as we showed to be the case for small Josephson energy,  signatures of the underlying complex internal dynamics are encoded in rare trajectories, which in the $s$-ensemble correspond to dynamical phases away from $s=0$.  Thus we demonstrated how the $s$-ensemble provides a method for interpretation of accurate measurements of the distribution of trajectories.

Exploring the SSET-resonator system also allowed us to examine trajectories in the $s$-ensemble created from different operators.  As each of the coupled components is itself an open system, we were able to study both trajectories formed by measuring charges in the right-hand SSET lead and photons emitted by the resonator.  Previous studies have shown that fast oscillators imprint a signature of their energy absorption from the SSET on the SSET charge statistics~\cite{Harvey2008}.  We demonstrated that, in this regime, the phase structure of trajectories of decaying quasiparticles mirrors that found by measuring photons emitted by the resonator.  

Finally, we emphasize the potential significance of this work with regard to future experiments.  SSET-resonator systems have been constructed with a stripline resonator~\cite{Astafiev2007}.  Impressive progress towards measuring single microwave photons~\cite{Bozyigit2011,Eichler2011,Zakka-Bajjani2010,Chen2011} makes measuring the trajectories studied in this work an exciting possibility.

\begin{acknowledgements}

We wish to thank P.~Kirton for useful discussions.  We gratefully acknowledge financial support from The Leverhulme Trust under grant no. F/00114/B6.

\end{acknowledgements}

\appendix*

\section{Practical Requirements}

In this Appendix we give a brief overview of the practical requirements for reaching a regime where the dynamics of a SSET-resonator system is described by Eqs.~\eqref{eq:ham} to~\eqref{eq:Losc}. Fuller accounts are available in works which focus on the theory of SSETs~\cite{Clerk2002, Choi2001} or SSET-resonator systems~\cite{Choi2003, Clerk2005, Blencowe2005, Rodrigues2007a}  and in experimental studies of SSETs coupled to either a mechanical~\cite{Naik2006} or electrical resonator~\cite{Astafiev2007}.

We start by considering the conditions that must be fulfilled for the current through the SSET to be dominated by JQP processes, which are composed of cycles involving a combination of coherent Cooper-pair tunneling and quasiparticle decay. The JQP resonances occur in the low temperature regime where the electrostatic charging energy of the SSET island, $E_c=e^2/2C_{\Sigma}$, with $C_{\Sigma}$ the total capacitance of the island, is much larger than $k_{\rm B}T$. Within this regime, the number of different charge states accessible to the SSET island becomes severely limited and we can treat it as a few-level system. The other two important energy scales for the SSET are the superconducting gap, $\Delta$, and the Josephson energy associated with the junctions $E_J=h\Delta/(8e^2R_J)$, where $R_J$ is the resistance of the Josephson junctions.

A JQP resonance occurs when the gate and bias voltages, $V_g$ and $V_{ds}$ respectively, are chosen so that the electrostatic energy difference between states on the island differing by one Cooper-pair, $\Delta E$, is zero and the drain-source voltage is large enough so that both of the quasiparticle tunneling processes associated with the cycle are allowed. The gate voltage induces a polarization charge $n_g=(C_gV_g+C_JV_{ds})/e$ on the island~\cite{Blencowe2005}, where $C_g$ and $C_J$ are the capacitances of the gate and the Josephson junctions respectively. This then leads to an electrostatic energy difference $\Delta E=-4E_c(n_g-n-1)$  where $n$ is an integer, corresponding to the number of Cooper-pairs on the island,  (we take $n=0$ for simplicity in the main text) and hence resonance occurs along lines in the $V_g-V_{ds}$ plane given by $n_g=n+1$~\cite{Blencowe2005}. Both of the quasiparticle processes involved in the JQP cycle are allowed for $eV_{ds}>2\Delta+E_c$; provided the charging energy is large enough compared to the gap, $E_c>2\Delta/3$, the JQP cycle dominates the current up to  $eV_{ds}=4\Delta$, at which point current can flow through the motion of quasiparticles alone. The master equation description of the quasiparticle tunneling processes is derived assuming high resistances for the junctions~\cite{Clerk2002, Blencowe2005, Clerk2005}, $R_J\gg h/e^2$. 

The coupling of a nanomechanical resonator to a SSET is discussed in detail in~\cite{Clerk2005, Blencowe2005}, while the case of coupling to a superconducting stripline resonator is discussed in~\cite{Astafiev2007}, though from a theoretical point of view the form of the coupling is essentially the same in both cases.  In the case of a mechanical resonator, a metal layer is added to a nanomechanical beam fabricated from a semiconductor which is adjacent to the SSET island. The beam acts as an additional voltage gate and the length-scale, $x_s$, which describes the coupling between the island and the resonator [see Eq.~\eqref{eq:ham}] is given by
$x_s=2E_cC_mV_m/(em\omega^2d)$, where $C_m$ is the beam-island capacitance, $V_m$ is the voltage applied to the beam, $d$ is the distance between the beam and the island and $m$ the effective mass of the resonator. Temperatures $\sim 30$\,mK~\cite{Naik2006} are routinely used in experiments with SSETs.  We note that experiments using a superconducting stripline resonator have the advantage that the relevant mode frequency of the stripline, $\omega$ are large enough so that $\hbar\omega\gg k_{\rm B}T$ under typical conditions. For example, in the experiment using a stripline resonator reported in Ref.~\cite{Astafiev2007} the mode frequency was $\omega/2\pi=9.9$\,GHz.

\vspace{-1.4\baselineskip}


%

\end{document}